\documentstyle[preprint,aps,prl]{revtex}
\begin{document}
\textwidth 16cm
\title {\bf Carbon superatom thin films}

%
%
\author{Andrew Canning$^{(a,b)}$, Giulia Galli$^{(b,*)}$ and
Jeongnim Kim$^{(c)}$ }

\address{
$(a)$~Cray Research, PSE, EPFL, 1015 Lausanne, Switzerland.}
\address{
$(b)$~Institut Romand de Recherche Num\'{e}rique
en Physique des Mat\'{e}riaux (IRRMA),
IN-Ecublens, 1015 Lausanne, Switzerland.}
\address{
$(c)$~Department of Physics, The Ohio State University,
Columbus OH 43210, USA.$^{\dag}$}
\maketitle
%
%
%
\begin{abstract}
 Assembling clusters on surfaces has emerged as a novel way
to grow thin films with targeted properties.
In particular, it has been proposed from experimental findings that
fullerenes deposited on surfaces
could give rise to thin films retaining the 
bonding properties of the incident clusters. However
the microscopic structure of such films is still unclear.
By performing quantum molecular dynamics simulations,
we show that C$_{\rm 28}$ fullerenes can be deposited on a surface
to form a thin film of nearly defect free molecules, which act as
carbon superatoms.
Our findings help clarify the structure of disordered small
fullerene films and also support the recently proposed 
hyperdiamond model for solid C$_{\rm 28}$.
\end{abstract}
\newpage
A growing community of physicists, chemists
and material scientists have recently devoted a significant effort to
studying and controlling the behaviour of small clusters,
the main goal being the synthesis of new materials\cite{sciencerev1}.
In particular, assembling clusters by deposition on surfaces is
emerging as a promising way to 
grow thin films with targeted properties.
Since the experimental discovery of C$_{\rm 60}$\cite{KHOCS85},
many investigations have focussed on fullerenes,
due to the variety of exciting properties of
pure and doped systems in both the gas and solid phases. 
 
 Recent experiments\cite{PMDPPC93} suggest that low energy deposition of
fullerenes on semiconducting substrates can produce thin 
films which retain the bonding properties of the incident
clusters. This memory effect might permit the synthesis of amorphous
diamond or graphitic-like films depending on whether small
(C$_{\rm n}$, n=20-32) or large (C$_{\rm 900}$) clusters are  
deposited\cite{PMDPPC93}.
 Of all the small fullerenes, C$_{\rm 28}$ is the most studied:
this molecule is believed to be the smallest fullerene created
in the photofragmentation
\cite{AWMSS93} of C$_{\rm 60}$ by loss of C$_{\rm 2}$,
as well as being the smallest fullerene produced in considerable amounts
in the laser vapourisation of graphite\cite{GDCAHMOWSS92}.
The structure of the isolated cluster was first proposed by Kroto\cite{K87},
who suggested that C$_{\rm 28}$ should be 
chemically very reactive, with four
preferred active sites.
Recently, the existence of solid forms of C$_{\rm 28}$  has been conjectured. 
Guo et al.\cite{GDCAHMOWSS92} proposed that it should be possible to stabilise
an arrangement of the fullerene cages in
a diamond lattice (hyperdiamond structure).
Since then several studies of various ordered C$_{\rm 28}$ solids 
have appeared in the literature\cite{ASBK93,BK93,KZAJ94,ZJKA95},
however, the possible growth and stability of C$_{\rm 28}$ solids as well as
the ability of the clusters to form thin films remain at present open
questions.
 
\vskip 0.1 cm 

 In order to address these issues,  we performed
quantum molecular dynamics simulations to investigate the deposition process
of C$_{\rm 28}$ cages on a non-metallic substrate, corresponding to the early
stages of the growth of a disordered thin film. 
We used a tight-binding model\cite{QWC94} in the context of
a linear scaling method\cite{MG94,KMG95,AGMDC96}. 
Here we present an analysis of the deposition process, we discuss the
microscopic structural properties of the deposited film and relate them
to recent experimental findings\cite{PMDPPC93,GDCAHMOWSS92}.

 In our simulations, the interatomic potential is derived from the
total energy of interacting ions and electrons, as described by a 
tight-binding (TB) model\cite{QWC94}. 
This model has been successfully used to study a variety of
carbon systems\cite{QWC94}. We chose a TB picture rather than a classical
potential since the high reactivity of C$_{\rm 28}$ clusters and their 
ability to form
covalent bonds, call for a quantum mechanical treatment of the
interactions. As we will justify, a TB picture is appropriate for the
purpose of this study. 
We performed simulations with cells of up to 4472 atoms, for
several tens of picoseconds. These calculations of unprecedented size
were made possible by the favourable scaling
with system size of the algorithm that we adopted, together
with its efficient implementation on a massively parallel computer\cite{AGMDC96}.
 
 The validity and accuracy of linear scaling [O($N$)] methods 
for systems similar to those studied in the present work
have been addressed
in several recent publications\cite{QWC94,MG94,KMG95,O9596}.
In order to test the validity of our
O(N)-TB model for the description of C$_{\rm 28}$ systems, we compared 
results\cite{ftn11} for the C$_{\rm 28}$ monomer, dimer and hyperdiamond
ordered solid with those of first principles local density (LDA)
functional calculations. 
The comparison is reported in the Table and shows that 
the agreement between LDA and O(N)-TB results is extremely good. 
We therefore conclude
that the O(N)-TB approach is a sufficiently accurate model
to describe  inter- and intra-C$_{\rm 28}$  interactions.

 In our computer experiment\cite{ftn11,ftn12}, we projected 78 fullerenes 
onto a diamond substrate, of which 50 remained to form a film (see Fig.~1).
The substrate consisted of 
12 C layers with 256 atoms per layer with periodic boundary
conditions in the $x$ and $y$ (horizontal) directions, and a 
(2 $\times$ 1) reconstructed (111) surface on the upper and lower side.
The deposition was performed by sending down 14 separate
showers of C$_{\rm 28}$ clusters: the initial configuration of
each shower was prepared by choosing molecules 
randomly distributed in space and
noninteracting with each other and with the surface 
(due to the localisation of the orbitals). 
They were also chosen to have random orientations.
The number of fullerenes belonging to each shower varied from 4 to 8
depending on the maximum number of non-interacting molecules  whose center
of mass could be randomly placed on a plane parallel to, and of the
same size as the surface layer.
With a series of test runs (involving the study of fullerene-fullerene
and fullerene-surface impacts as a function of energy),
we determined the range of incident kinetic energies (20-40 eV)
for which the fullerenes had the highest sticking probability without breaking up.
We then projected each shower, one after the other,
vertically downwards onto the surface,
with each molecule having an initial kinetic energy randomly
distributed within this range of energies.  In this way we tried to mimic
some of the features of low-energy neutral-cluster beam deposition (LECBD)
experiments\cite{PMDPPC93}.
The bottom three layers of the substrate were held fixed and the next three layers
were connected to a Nos\'{e}-Hoover thermostat\cite{N84W85} 
at room temperature to prevent the substrate heating up\cite{WWC92} under the
many fullerene impacts.
After each shower of fullerenes the whole system
was equilibrated, by monitoring  
the variation in time of the number of bonds as well as the bond lengths.
 
At the end of the deposition process,
24 fullerenes were bonded directly to the substrate,
with the majority of them sticking to the surface during the first 3-4
showers. The arrangement of the first monolayer
of C$_{\rm 28}$'s on the surface is random, with no
preferential bonding sites (see Fig.~2a).
As expected from the large number of dangling bonds in the cluster,
C$_{\rm 28}$'s easily attach to the diamond substrate,
the average number of bonds with the surface being about three per 
fullerene.
These can be classified as strong covalent bonds: 
their formation can cause bonds between the fourfold coordinated atoms of the
first and second surface bilayer to break.
This locally destroys the
(2 $\times$ 1) Pandey chain reconstruction, returning the
surface geometry to an ideal (111) configuration (de-reconstruction).
Nevertheless the formation of bonds between
the substrate and C$_{\rm 28}$'s does not induce
modifications of their molecular topology. 
This is at variance to what was observed
for C$_{\rm 60}$ on a diamond (111) surface\cite{GM94}, 
where impacts leading to bonds between the C$_{\rm 60}$ 
and the surface would necessarily induce the formation
of defects in the cage, which could lead to its break up in the
high energy regime\cite{GM94}. 

 On the time scale of our simulation we did
not observe any diffusion of the C$_{\rm 28}$'s on the surface. 
Most importantly, when heating the system after the deposition was completed,
no diffusion was observed before reaching the temperature
regime where cages started to break up.
Therefore, we expect that the bonds between the fullerenes and the
surface are strong enough to hamper rearrangements and diffusion.
Consequently, the  system is unlikely to undergo a disorder to order 
transition like the one observed experimentally
in the case of C$_{\rm 60}$ on some semiconducting substrates\cite{XCC93}.
Indeed, for C$_{\rm 60}$, van der Waals interactions 
between the molecules and the substrate play a 
crucial role in allowing cluster diffusion and the 
subsequent disorder to order transition.

 On top of the first monolayer, a second portion of the system 
(26 fullerenes) was identified, which was again very inhomogeneous.
In this portion of the system, during the deposition process, 
we observed the formation of 
three-dimensional islands with some polymer-like structures, where
C$_{\rm 28}$'s were preferentially bonded to each other
with two single bonds (see Fig.~2b and 2c). 
This is suggestive of either a Stranski-Krastanov or a
Volmer-Weber growth process\cite{book}. 
However our simulation cannot clearly establish a growth process mechanism since
larger cell sizes and longer simulation times would be needed\cite{ftn3}.
 After equilibration at room temperature, the film was heated up 
to verify its thermal stability.
During the heating procedure, we observed oscillations of the polymer-like 
structures with no major structural change. 
Our results indicate that the film is stable up to about 1000 K\cite{ftn3}.

The whole system at the end of the deposition process at room
temperature is shown in Fig.~1.
The thin film is characterised by fullerenes which retained their topology,
with small distortions\cite{ftn2} with respect to the original shapes.
These results are in agreement with the memory effect reported experimentally
when depositing small fullerenes by LECBD\cite{PMDPPC93}.
Furthermore in our deposited film the particle density is very low 
($\simeq$ 1 gr cm $^{-3}$, compared
to the diamond and graphite densities of 3.52 and 2.27 gr cm$^{-3}$, respectively)
consistent with the measured value of 0.8 $\pm$ 0.2 gr cm$^{-3}$ for
small fullerene films obtained by LECBD\cite{PMDPPC93}.

 We found that the total particle-particle correlation function
of the deposited film
has two main peaks corresponding to first and second neighbour
distances. Each of these peaks is split, with the first
double peak at 1.42 and 1.53 \AA: these distances correspond to 
the two different intra-fullerene bond lengths (see Table). 
The longer distance
also corresponds to the inter-fullerene bond length: the
correlation function of the sites involved in inter-molecular bonds
has two single main peaks, the first being centered at 1.52 \AA. 
Most fullerenes are three- and 
fourfold coordinated,  although some two-fold and five-fold cages
were also observed. 
This predominance of three- and fourfold coordinated molecules indicates
that the C$_{\rm 28}$ cluster has bonding properties similar to
those of C atoms in amorphous systems, which usually 
contain a mixture of sp$^2$ and sp$^3$ sites\cite{R91}. 
 This is in agreement with the proposition of Kroto\cite{K87}
that C$_{\rm 28}$ should behave as a carbon superatom with four
preferred active sites (A sites; see Table for definition of site types).
 We found that A sites have the highest probability (0.37)
to form bonds, giving rise to hyperdiamond-like configurations. 
They are then followed by C (0.11) and B (0.04) sites, 
in order of decreasing probability.
We note that the structural properties of the film deposited
in our simulation are remarkably similar to those of
a disordered system obtained by subsequent collisions of C$_{\rm 28}$'s,
mimicking gas phase deposition of the clusters\cite{KGWC96}.
Our findings support
early conjectures of Guo et al\cite{GDCAHMOWSS92} 
that C$_{\rm 28}$ solids in a hyperdiamond
structure are stable, and are in agreement with the study of
Kaxiras et al.\cite{KZAJ94} on the reactivity of C$_{\rm 28}$.

 The proportion of fourfold coordinated atoms and thus of
proper sp$^3$ sites is low  (10 $\%$) in the deposited film.
Experimentally, from analyses of Raman spectra 
it has been suggested\cite{PMDPPC93} that 
films obtained by depositing small fullerenes might be composed of  
small graphitic islands embedded in a diamond-like matrix.
We do not find any evidence of such structures in our deposition energy regime.
However we note that the computed vibrational spectra (both within LDA and TB)
of free C$_{\rm 28}$\cite{OGK96} (and C$_{\rm 20}$ \cite{GGG}) cages 
are consistent with the reduced Raman spectra calculated from experimental data 
on small fullerene films. Neither type of spectra exhibit  prominent peaks
corresponding to graphite Raman active modes (the
maximum vibrational frequency\cite{OGK96} 
for the fullerene cages is about 1500 cm$^{-1}$)
 and  both spectra show peaks centered around
600 and 1200 cm$^{-1}$. This indicates that the experimentally 
measured Raman spectra
might show direct evidence for the presence of cages in the film,
consistent with the observed memory effects. 
Calculation of the phonon density of states 
of the deposited film is underway and will help
clarifying this point\cite{OGK96}.

 In conclusion, 
our work shows that using O(N) methods it is now possible, 
for certain types of system, to perform TB molecular dynamics simulations
on supercells of several thousand atoms for tens of picoseconds.
This allows phenomena, previously inaccessible to quantum molecular
dynamics, to be studied.
In the O(N)-TB molecular dynamics simulations presented in this paper,
we have shown that in a given energy range
small fullerenes can be assembled on a semiconducting surface
to form the first layers of a thin film where the C$_{\rm 28}$'s constitute 
the building blocks and act as carbon superatoms, giving rise
to hyperdiamond-like configurations. Our results are in agreement with
memory effects detected in small fullerene films deposited by 
LECBD\cite{PMDPPC93} and help clarify the structure of
disordered small fullerene films. They also suggest the possibility to 
synthesize
novel new thin films, taking advantage of the many and varied properties of
pure and doped fullerene cages.

\newpage


%
\newpage
\begin{table} 
\begin{tabular}{|lcc|}  
{\bf C$_{\bf 28}$ monomer  }  & O(N)-TB & LDA \\
\hline 
 A-B bond length (\AA )  & 1.44-1.46& 1.42-1.45  \\
 A-C bond length (\AA ) & 1.40-1.44& 1.40-1.44  \\
 C-C bond length (\AA ) & 1.47-1.55& 1.46-1.57  \\
\hline 
{\bf C$_{\bf 28}$ dimer } & O(N)-TB & LDA \\
\hline 
A-A bonded {\bf R$_{\bf eq}$} (\AA ) & 1.54  &  1.51          \\
A-C bonded {\bf R$_{\bf eq}$} (\AA )& 1.46  &  1.51          \\
C-C bonded {\bf R$_{\bf eq}$} (\AA )& 1.50  &  1.53          \\
A-A bonded {\bf E$_{\bf coh}$} (eV) & 3.02    & 3.15           \\
A-C bonded {\bf E$_{\bf coh}$} (eV) & 2.13    & 2.48           \\
C-C bonded {\bf E$_{\bf coh}$} (eV) & 2.24    & 2.48         \\
\hline 
{\bf Hyperdiamond} & O(N)-TB & LDA \\
\hline 
Lattice constant (\AA )& 15.85 & 15.78 \\
{\bf E$_{\bf coh}$}(eV) & 0.65 & 0.74 \\
C$_{\rm 28}$-C$_{\rm 28}$ bond length (\AA )& 1.52 & 1.54 \\
\end{tabular}
\vskip 0.7cm
\end{table}
TABLE CAPTION\\
\noindent
Comparison between properties of the C$_{\rm 28}$ monomer and dimer and of
hyperdiamond obtained from first-principles calculations using the
local density functional theory (LDA), and
from semi-empirical tight-binding (TB) calculations using the O($N$) method of
Ref.~\protect\cite{KMG95}. The LDA calculations for hyperdiamond are taken from
Ref.~\protect\cite{KZAJ94}. The LDA calculations for the monomer and the dimer
were carried out within the 
pseudopotential-plane-wave formalism\cite{CP85}, using 
Troullier-Martins pseudopotentials\protect\cite{TM91} and  
a kinetic energy cutoff of 30 and 44 Ry (to check for convergence).
We used supercells of dimensions (15.9 \AA\ )$^3$ and
(15.9 \AA $\times$ 15.9 \AA $\times$ 23.8 \AA) for the monomer and
the dimer, respectively. Since the C$_{\rm 28}$ molecule is Jahn-Teller
distorted and polarisation effects can be important, we checked for
convergence of total energy with respect to cell size
and found that in a cell of (15.9 \AA\ )$^3$ the energy was converged up 
to 0.005 eV.
The C$_{\rm 28}$ monomer
is composed of 12 pentagons and 4 hexagons; 
atoms at an apex where three pentagons meet are denoted as A sites; 
atoms on the hexagons are denoted as B sites if bonded to A sites, and C 
sites otherwise. The C$_{\rm 28}$ molecule has 12 A-B, 24 B-C 
and 6 C-C bonds.
The C$_{\rm 28}$ dimer is composed of two C$_{\rm 28}$ molecules
connected by a single bond with equilibrium bond length ${\bf R_{\bf eq}}$: 
we considered three cases involving A and C atoms as bonding sites, since
these are the most reactive sites (see text and 
Ref.~\protect\cite{K87,ASBK93,BK93,KZAJ94}).
Ionic positions of both the monomer and the dimer were fully optimised
and the equilibrium rotational configuration of the dimer was the same for
both O(N)-TB and LDA calculations.
The cohesive energy of the dimer was taken to be 
${\bf E_{\bf coh}} = (E^{\rm D} - 2 \cdot E^{\rm M})$, where $E^{\rm D}$ and 
$E^{\rm M}$ are the total energies of the dimer and the monomer, respectively.
The cohesive energy of hyperdiamond is with respect to the cohesive
energy of diamond. 
In the O($N$)-TB calculation for hyperdiamond
both the lattice constant  and the atomic
position were optimised, whereas in the LDA calculation only the relative
position of the C$_{\rm 28}$'s, assumed to be in a perfect tetrahedral geometry, 
was optimised\protect\cite{KZAJ94}. 
O($N$)-TB calculations with 8 and 64 C$_{\rm 28}$'s per cell
gave almost identical results for the properties of hyperdiamond reported in the
Table.
%
%
%
\newpage
\figure{ {\bf Fig.~1}
A snapshot of the full system (4472 atoms) at the end of the molecular dynamics
deposition simulation showing the undamaged C$_{\rm 28}$ cages.
Red and blue spheres represent atoms belonging to the surface
and to the fullerenes, respectively. Bonds between surface atoms
and between fullerene atoms are represented as light and dark green,
respectively.
Note the numerous bonds between the fullerene and the first surface
layer causing the local de-reconstruction of sections of the Pandey chains.
}
\figure{ {\bf Fig.~2}
Sections of the system displayed in Fig.~1 (colour scheme 
as defined in Fig.~1 legend).
(A) Top view of the top layer of the substrate showing only 
the fullerenes directly bonded to the surface:  
note the random positions of the fullerenes bonded to the surface.
(B) top and (C) side view of the top layer of the substrate showing only
the fullerenes which are {\em not} bonded to the surface:
the tendency of the fullerenes to form islands and polymer type 
structures is clearly visible.
}

\begin{thebibliography}{10}
\bibitem{ftn12}
In the deposition
simulation we performed  15 iterations per ionic
move to diagonalise the TB Hamiltonian,  
with a time step for the integration of the
equations of motion of 10 a.u.  The total simulation time was 70 ps.
\bibitem{ftn3}
 From our heating experiment, we concluded that the equilibration of
a film containing more that $\simeq$ 50 fullerenes would become extremely long and
eventually unaffordable, due to the low frequency modes associated with
the oscillations of the polymer-like structures.
\bibitem{ftn2} 
In the deposited film, the maximum and minimum ratio between 
the principal moment of inertia are 1.3 and 1.1, respectively and
the variations of the maximum and minimum distance within a 
cluster are 1.54-1.65 and 1.30-1.40 \AA\ , respectively.
\end{thebibliography}
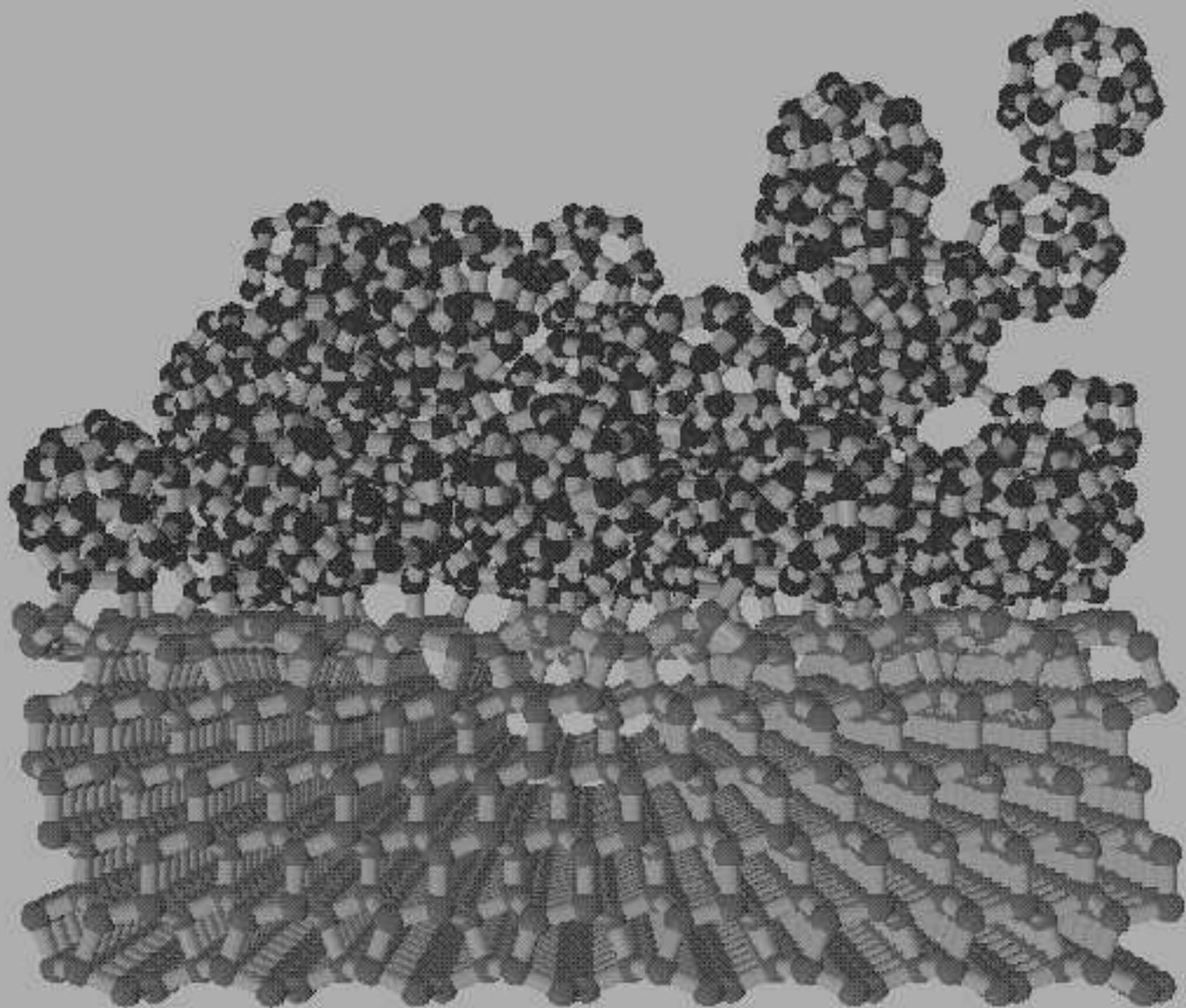
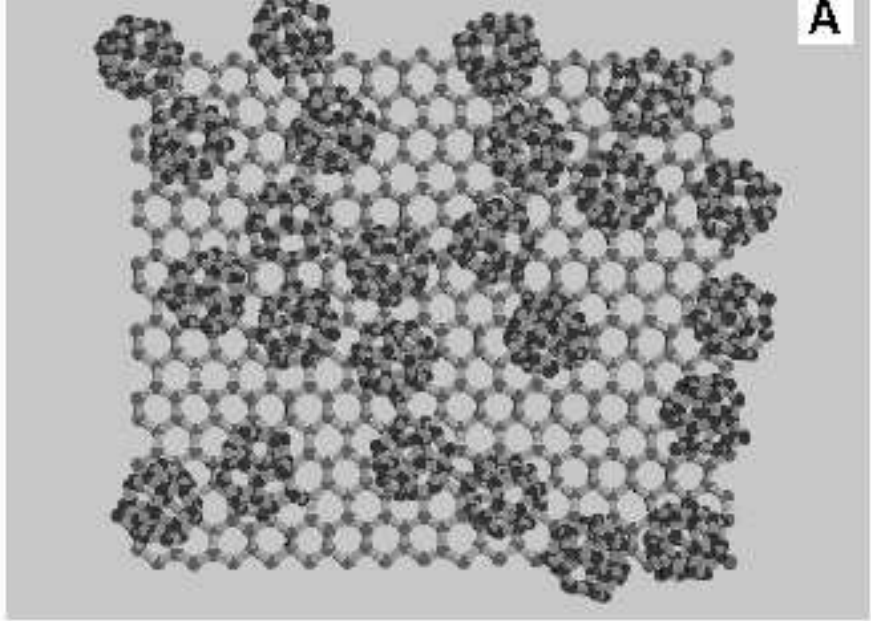
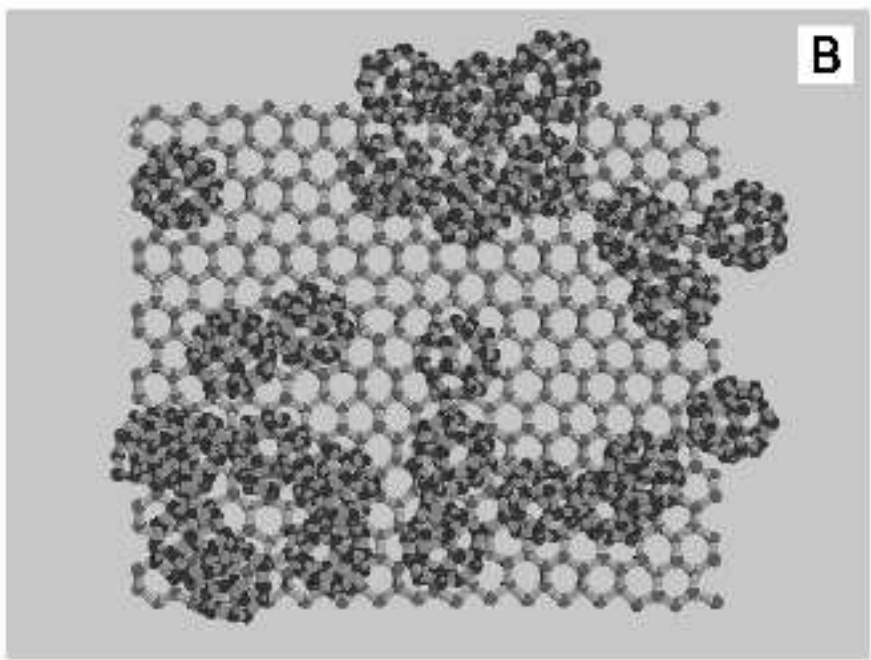
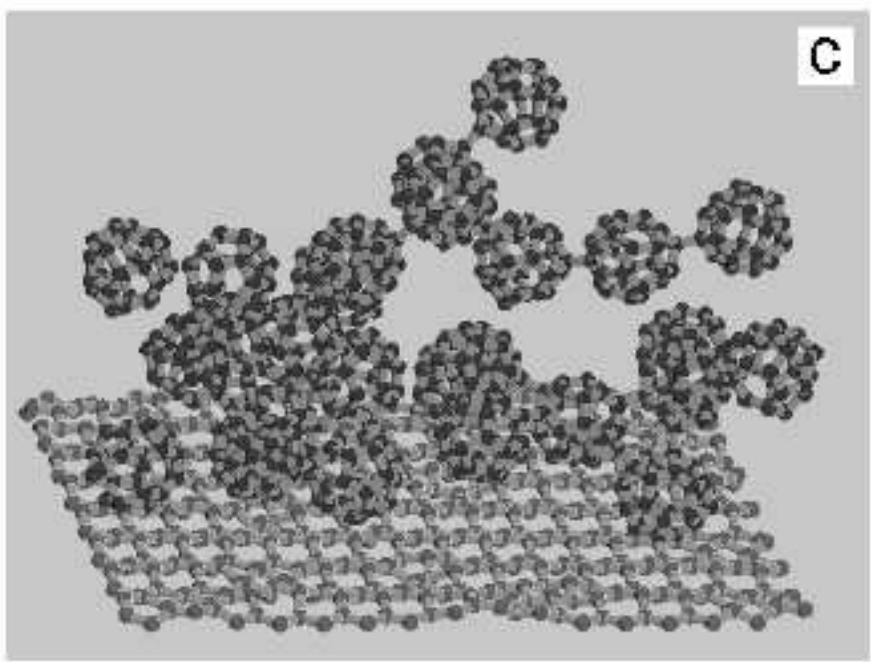
\end{document}